\documentclass[rn,11pt]{ucl_document} 
\usepackage{graphicx}
\usepackage{hyperref} 
\usepackage{named}

\newlength{\halfwidth}
\newlength{\halfwidthextra}
\begin{document}

\title{More Mouldy Data: Virtual Infection of the Human Genome}
\date{14 June 2011}
\documentnumber{11/14}

\contactdetails{%
\mbox{\emph{Electronic Mail}: W,Langdon@cs.ucl.ac.uk}\\%
\mbox{\emph{URL}: 
\href{http://www.cs.ucl.ac.uk/staff/W.Langdon/}
{http://www.cs.ucl.ac.uk/staff/W.Langdon/}}\\%
}
\author{W. B.~Langdon and M. J. Arno}
\maketitle


\begin{abstract}
The human genome sequence database contains 
DNA sequences very like those of \mbox{mycoplasma} molds.
It appears such moulds infect not only molecular Biology laboratories but
were picked up by experimenters from contaminated samples
and inserted into GenBank as if they were human.
At least one mouldy EST (Expressed Sequence Tag) has transferred from 
public databases to commercial tools
(Affymetrix HG-U133 plus 2.0 microarrays).
We report a second example (DA466599) and
suggest there is a need to clean up genomic databases
but fear current tools will be inadequate to catch genes
which have jumped the silicon barrier.
\end{abstract}

\keywords{Bioinformatics, data cleansing,
bit rot, {\em in silico} biology, Blast, Phishing,
jumping information genes
}

\section{Introduction}

Ensuring databases are both up to date and
contain only correct data is a huge software engineering problem.
Even as the human genome was first published
the associated problems of data cleansing 
Bioinformatics sequence data were being discussed
\cite{Genome_Res.-1999-Felsenfeld-1-4,%
Bioinformatics-2001-Apweiler-533-4-1}
but it appears only technical problems where considered.

We discovered that the definitive publicly accessible
database holding the human DNA sequence has been corrupted
in a surprising way.
It contains the DNA sequence of a mold \cite{Astarloa:2009:BT}.

More recently we have discovered a second sequence which is probably
not human in the human genome.
It appears that the time is ripe for a though check of the NCBI GenBank
database.

It appears that not only has the Human DNA sequence been
``completely sequenced'' \cite{Genome_Res.-1999-Felsenfeld-1-4}
but in the process other living organisms
commonly found in Molecular Biology laboratories
have infected not just the physical samples
but also the virtual {\em in~silico} Bioinformatics environment.
By unwittingly using a technique reminiscent of computer hacking,
a mold gene has succeeded in not just 
moving within its own genome~\cite{McClintock01031929}
nor only jumping horizontally 
and crossing the species barrier~\cite{Akiba:1960:JJMB}
but has crossed the silicon barrier between life and data
and succeeded in reproducing itself across very diverse information
based media.
Given the highly interconnected nature of genomic research, 
technology and medicine
and the low priority so far attached to the problem,
it is unlikely current data warehouse cleansing techniques will be
able to eradicate this and potentially other silicon jumping genes.

\section{Computational {\em in silico} Experiment}
\label{sec:results}

The anomalous HG-U133~+2 sequence
(GenBank AF241217, probeset 1570561\_at)
we had previously reported \cite{Astarloa:2009:BT}
was run against the human genome
using Blast~\cite{ncbi_blast}, at 
the European Bioinformatics Institute EMBL-EMI
with their default settings.
This gave a list of DNA sequences
which partially match published DNA sequences.
The list is ordered by blastn so that the best matches are at the top.
Only the top 50 fuzzy matches are included in the list.
As expected the first match is the query sequence itself
(EM\_HTG:AF241217).
Despite~\cite{Astarloa:2009:BT} having been published 
more than a year ago,
EM\_HTG:AF241217 is still described as ``Homo sapiens''. 
All the others are mycoplasma, except the 34$^{th}$ in the list, DA466599,
which EBI says is human.
(EBI gives one reference for DA466599:~\cite{Genome_Res.-2006-Kimura}.)
%
However we suggest that DA466599 may not be a human DNA sequences
but is another example of physical contamination
leading to virtual infection of the public data.

We ran a second EBI blastn query (again using the NCBI em\_rel database).
This time looking for DNA sequences that match DA466599. 
The results for DA466599 are similar to those for AF241217
and so support the view
that DNA sequence DA466599 is not human
but instead is also a contamination.
Again the best 50 matches were reported.
Of course the first one is DA466599 itself.
All the other matches returned by blastn
are for various species of mycoplasma.

\section{Discussion}
\label{sec:discuss}


It is well known that mycoplasma contamination is rife in 
molecular biology laboratories \cite{Miller:2003:BioTech}.
Many labs are routinely periodically sterilised to counter it.
Miller {\em et al.}~\cite{Miller:2003:BioTech}
said mycoplasma contamination has 
``potentially major consequences for the diagnosis and
characterization of diseases using expression array technology.''
Nonetheless,
using RNAnet\footnote{
\url{http://bioinformatics.essex.ac.uk/users/wlangdon/rnanet/}},
we previously estimated about 1\% of published data
in the
Gene Expression Omnibus (GEO) database at NCBI
(\url{www.ncbi.nlm.nih.gov/geo})
are contaminated \cite{Astarloa:2009:BT}.

One potential fortuitous side effect of the 
{\em in~silico} spread of mycoplasma contamination is that the
Affymetrix HF-U133~+2 1570561\_at probeset
might be used to indicate physical sample contamination.
Thus probeset 1570561\_at could be treated as a free additional
quality control signal.
If 1570561\_at says there is significant expression of mycoplasma
genes, then the sample is probably contaminated and
the other gene expression levels given by the microarray are suspect.

Having found two suspect DNA sequences it seems likely 
the published ``human genome'' sequence contains more.
Indeed contamination of all organism sequences seems possible. 
With the explosive growth of genomic sequence data available via the
Internet,
including data from the 1000 genome project \cite{nature09534},
it seems time to look again at genomic database quality.




\section*{Acknowledgments}

Matthew Arno manages the Genomics Centre, 
King's College London, 
matthew.arno@kcl.ac.uk.


EPSRC grant
\href{http://gow.epsrc.ac.uk/NGBOViewGrant.aspx?GrantRef=EP/I033688/1}
{EP/I033688/1}.


\bibliographystyle{unsrt} 
\bibliography{references}

\end{document}